\documentclass[
 aip,
 amsmath,amssymb,
 reprint,
]{revtex4-1}

\usepackage{graphicx}
\usepackage{dcolumn}
\usepackage{bm}

\usepackage[utf8]{inputenc}
\usepackage[T1]{fontenc}
\usepackage{mathptmx}
\usepackage{etoolbox}

\usepackage{booktabs}

\makeatletter
\def\@email#1#2{%
 \endgroup
 \patchcmd{\titleblock@produce}
  {\frontmatter@RRAPformat}
  {\frontmatter@RRAPformat{\produce@RRAP{*#1\href{mailto:#2}{#2}}}\frontmatter@RRAPformat}
  {}{}
}%
\makeatother
\begin{document}

\title[Machine learning methods for modelling local, linear gyrokinetic
simulations of MAST-U pedestal turbulence]{Machine learning methods for modelling local, linear gyrokinetic
simulations of MAST-U pedestal turbulence}
\author{A. Niemelä}
 \email{anna.niemela@vtt.fi}

\author{D. Jordan}%
\affiliation{ 
VTT Technical Research Centre of Finland, Espoo, Finland
}%

\author{A. Järvinen}
\affiliation{
VTT Technical Research Centre of Finland, Espoo, Finland
}%

\author{A. Bruncrona}
\affiliation{
VTT Technical Research Centre of Finland, Espoo, Finland
}%

\author{A. Kit}
\affiliation{
VTT Technical Research Centre of Finland, Espoo, Finland
}%

\author{L. Frassinetti}
\affiliation{
Department of Electromagnetics and Plasma Physics, KTH Royal Institute of Technology, Stockholm, Sweden
}%

\author{D. Hatch}
\affiliation{
Institute for Fusion Studies, University of Texas at Austin, Austin, USA
}%

\author{L. Leppin}
\affiliation{
Institute for Fusion Studies, University of Texas at Austin, Austin, USA
}%

\author{S. Saarelma}
\affiliation{
UK Atomic Energy Authority, Abingdon, UK
}%

\author{the MAST Upgrade team}
\affiliation{
See the author list of J. Harrison et al 2026 Nucl. Fusion 66 116005
}%

\author{EUROfusion Tokamak Exploitation team}
\affiliation{
See the author list of N. Vianello et al 2026 Nucl. Fusion 66 116010
}%

\date{\today}

\begin{abstract}
Gyrokinetic (GK) stability strongly influences the performance of high-confinement-mode pedestals in spherical tokamak plasmas. High-fidelity gyrokinetic codes such as GENE can model microinstability-driven transport, but the computational cost limits their routine use in integrated pedestal modeling workflows. Instead, present workflows often rely on reduced transport assumptions, such as the ballooning-critical pedestal model used in EPED.
This work investigates machine-learning surrogate models for local linear gyrokinetic simulations in a MAST-U-relevant pedestal parameter space, with the aim of providing faster gyrokinetic-based inputs to reduced pedestal models. A sampling workflow is developed in which pedestal profile parameters are varied within experimentally motivated bounds and used to generate physically self-consistent Grad--Shafranov equilibria. This reduces the dimensionality of the data-generation problem compared with sampling local gyrokinetic inputs directly, while maintaining physically plausible combinations of plasma profiles, geometry, and local stability parameters.
The surrogate models are trained to predict linear growth rates, real frequencies, and diffusivity-ratio transport fingerprints from local linear GENE simulations. A multi-head multilayer perceptron accurately reproduces the growth rate, while the diffusivity ratios and real frequency exhibit more clustered, regime-dependent behavior. A multi-head classification--regression model using frequency-based regime classes reduces the mean absolute error for these clustered targets and better captures sharp transitions associated with changes in the underlying instability regime, although errors near mode-transition regions remain a limitation. 
\end{abstract}

\maketitle



\section{\label{sec:intro}Introduction}

In high confinement (H-mode) tokamak plasmas, a self-organized transport barrier and a pressure pedestal forms at the edge of the plasma \cite{Wagner2007,Groebner2018}. The pedestal top acts as a boundary condition for the core plasma profiles, and therefore the pedestal height and width have a strong influence on the confined pressure and fusion performance \cite{Snyder2011EPED, Groebner2018}. 

A central challenge for future fusion reactors is whether a sufficiently high-performing pedestal can be achieved simultaneously with the power exhaust measures necessary to protect reactor components \cite{Leonard2014}. However,  due to the multiple physical processes and spatial scales, predicting this core-edge integration is very challenging, and the current approaches for estimating the scalings of pedestal performance between scenarios and devices are primarily conducted via reduced models \cite{Snyder2011EPED}. The present paradigm for predicting pedestal top plasma conditions is based on a linear evaluation of the overall MHD stability envelope coupled with a reduced transport model constraining the profile \cite{Snyder2011EPED}.

A limitation of these models is the reduced transport assumption, which for example in the predictive pedestal model EPED, is the ballooning critical pedestal approach, which assumes the pedestal is limited by the onset of kinetic ballooning modes (KBMs) \cite{Snyder2011EPED, Parisi2024KBM}. Although this approach has been successful in reproducing broad pedestal trends, comparisons with experimental pedestal databases show remaining discrepancies, to which the simplified treatment of transport may contribute \cite{Saarelma2019, Frassinetti2021}. Accurate transport models are also needed for ELM-free and ELM-suppressed scenarios, where pedestal evolution is expected to depend on continuous transport rather than transient ELM relaxation \cite{Viezzer2018ELMFree, Parisi2024KBM}. This motivates the development of models that include more direct information from gyrokinetic stability and transport calculations.

In practice, several microinstabilities, including ion and electron temperature gradient modes (ITG, ETG), trapped electron modes (TEM), micro-tearing modes (MTM) and KBMs, are expected to contribute to transport \cite{Kotschenreuther2019}. Gyrokinetic simulations can be used to model this microturbulence-driven transport in magnetized plasmas. Codes such as GENE solve the gyrokinetic Vlasov--Maxwell system and can compute instability properties and turbulent fluxes driven by microinstabilities \cite{Jenko2000}. Nonlinear gyrokinetic simulations provide a high-fidelity description of turbulence-driven transport, but their computational cost makes them impractical for routine use in integrated modeling, scenario optimization, or large parameter scans \cite{Pueschel2012,Hatch2016MTM}.

Quasilinear transport models provide one route to reducing this cost by approximating nonlinear turbulent fluxes using a spectrum of linear gyrokinetic modes together with a saturation rule \cite{Bourdelle2016QuaLiKiz}. Although substantially cheaper than nonlinear simulations, performing the large number of linear gyrokinetic calculations required for quasilinear modeling remains computationally demanding. This limits their practicality for applications such as real-time analysis, uncertainty quantification, broad parameter scans, and transport simulations for the full discharge.

The primary aim of this work is to investigate methods to develop machine-learning (ML) surrogate models for linear gyrokinetics in the pedestal. When applied in suitable quasilinear transport models, such surrogates could enable transport predictions with evaluation speeds comparable to present-day reduced models, while retaining a closer connection to the gyrokinetic description of the underlying plasma \cite{HornsbyMTM}. This is of particular interest for spherical tokamaks such as MAST Upgrade (MAST-U), where pedestal stability and transport scalings may differ from conventional-aspect-ratio tokamaks \cite{Parisi2024KBM}. In this work, we train surrogate models on local linear GENE simulations generated within a MAST-U pedestal parameter space based on a reference discharge. The models predict linear growth rates, real frequencies, and diffusivity-ratio fingerprints across radial position and wavenumber, with particular attention to regime-dependent behavior associated with changes in the dominant instability.

\section{Dataset generation}

\subsection{Physics background}

The motivation of this work is the use of linear gyrokinetic information in reduced transport models. In quasilinear approaches, turbulent transport is estimated from the properties of linear gyrokinetic modes, while the nonlinear fluctuation level is supplied by a model-dependent saturation rule \cite{Stephens2021,Staebler2024}. For example, in a schematic mixing-length form, the heat diffusivity of species $s$ may be expressed as
\begin{equation}
\chi_s \sim \sum_{k_y}
\mathcal{W}_s(k_y)
\frac{\gamma(k_y)}
{\left\langle k_\perp(k_y)^2 \right\rangle}
\end{equation}
where $k_y$ is the binormal wavenumber of the mode, $\gamma(k_y)$ is the linear growth rate, $\left\langle k_\perp^2(k_y) \right\rangle$ is a characteristic mode-averaged perpendicular wavenumber, and $\mathcal{W}s(k_y)$ is a model-dependent quasilinear weight describing how the corresponding linear mode contributes to the heat transport of the species \cite{Stephens2021,Xie2020}. The factor $\gamma/\langle k\perp^2\rangle$ represents a mixing-length estimate of the saturated fluctuation level, while the precise form of the weight and saturation prescription depends on the reduced transport model being used \cite{Staebler2024}.

Thus, quasilinear models require predictions of linear gyrokinetic quantities, across a range of wavenumbers and radial locations \cite{Stephens2021, Staebler2024}. Although linear simulations are significantly less expensive than nonlinear simulations, the need to evaluate them across many radial locations and wavenumbers still represents a substantial computational cost. This requirement becomes particularly restrictive in applications involving large parameter scans, uncertainty quantification, or real-time prediction.

To address this limitation, we construct a surrogate model trained on a dataset of linear gyrokinetic simulations. To accelerate the generation of data, the local flux-tube approximation is applied in this work, although it is acknowledged that the scale separation between the equilibrium profiles and turbulent fluctuations may not be fully justified in the ion-scale. Addressing this limitation will be considered in future studies. The present work focuses on surrogate prediction of a subset of the linear gyrokinetic quantities needed for this type of reduced modeling: the growth rate, real frequency, and diffusivity-ratio fingerprints, here $D_e/\chi_e$ and $\chi_i/\chi_e$. The growth rate sets the instability drive, the real frequency provides information about the mode propagation, and the diffusivity ratios describe the relative transport channels, which can be used in reduced models to relate particle, electron heat, and ion heat transport. Complete quasilinear transport models also require additional quantities, including the perpendicular wavenumber, together with a model for nonlinear saturation.

\begin{figure*}[t]
    \centering
    \includegraphics[width=\textwidth]{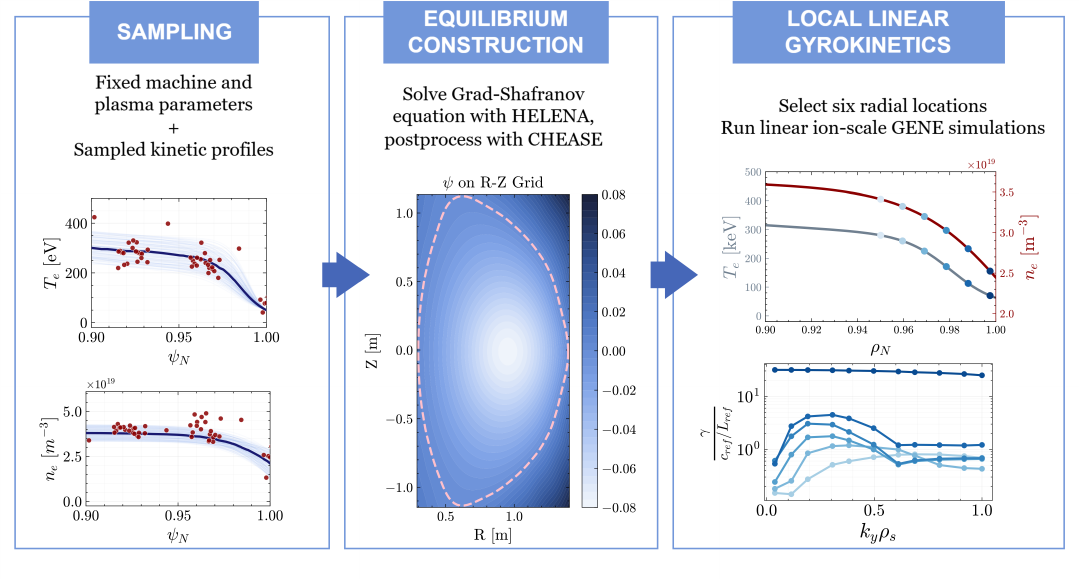}
    \caption{Workflow used to generate the training dataset for the surrogate model. 
Pedestal parameters are sampled within experimentally constrained bounds derived from a reference MAST-U discharge, and used to construct smooth temperature and density profiles via an $m$-tanh parameterisation. 
These profiles are input to the HELENA equilibrium solver to compute self-consistent pedestal equilibria, which are then used to define local flux surfaces within the pedestal region. At each selected radial location, linear gyrokinetic simulations are performed using GENE over a range of ion-scale binormal wavenumbers, producing growth rates, mode frequencies, and transport quantities that form the machine-learning training dataset.}
    \label{fig:workflow}
\end{figure*}

\subsection{Sampling}
For developing machine-learning models of gyrokinetic stability, a training dataset must span a physically relevant and representative portion of tokamak operational space. The large number of plasma parameters that can influence the stability, combined with the nonlinear nature of the underlying physics, requires a sampling strategy that efficiently covers the input space while maintaining experimental realism.

Rather than using experimental profiles directly, we use them to define sampling bounds, enabling the generation of physically plausible and generalizable inputs (Table \ref{tabI}). This approach supports scalable data generation and allows the surrogate model to be applied beyond existing experimental regimes in the future. The sampled parameters include pedestal top electron temperature and density, separatrix density, and the pedestal widths of both temperature and density profiles. The bounds for these parameters are derived from Thomson scattering measurements of MAST-U discharge 49108. This pulse was performed at 750kA/0.5T with 3.2MW of NBI heating. The pulse  is a type I ELMy H-mode that reaches a pedestal temperature around 0.3keV and in which the pedestal was found to be limited by ideal-MHD peeling instabilities \cite{Frassinetti2026}.

The Thomson scattering profiles are mapped onto normalized poloidal flux coordinates, and only data from time slices within 75\%–95\% of the inter-ELM period is used. This interval corresponds to phases in which the pedestal is nearly saturated and therefore representative of its quasi-steady-state structure. 
To ensure efficient coverage of the parameter space, the training dataset is generated using Sobol sequence sampling, which provides a quasi-uniform distribution of samples \cite{Sobol1967}.

\begin{table}[h!]
\centering
\caption{Sampled pedestal parameters and their ranges.}\label{tabI}
\begin{tabular}{l l c c}
\hline
\textbf{Parameter} & \textbf{Description} & \textbf{Units} & \textbf{Range} \\
\hline
$T_{e,\text{ped}}$   & Temperature, pedestal top      & keV                        & $[0.18,\,0.3]$ \\
$n_{e,\text{ped}}$   & Density, pedestal top          & $10^{19}\,\mathrm{m}^{-3}$ & $[3.3,\,4.2]$ \\
$n_{e,\text{sep}}$   & Density at separatrix          & $10^{19}\,\mathrm{m}^{-3}$ & $[1.3,\,2.9]$ \\
$\Delta_{T_e}$       & Temperature pedestal width     & $\psi$                     & $[0.02,\,0.04]$ \\
$\Delta_{n_e}$       & Density pedestal width         & $\psi$                     & $[0.05,\,0.07]$ \\
\hline
\end{tabular}
\end{table}

Sampling within these experimentally constrained bounds ensures that the generated profiles remain consistent with the observed variability of the reference discharge, as the ranges derived from Thomson scattering capture the experimental scatter in pedestal height and width.

As this work serves as a proof-of-principle focused on a single MAST-U discharge, quantities such as the equilibrium shape, toroidal field plasma current, species charges, and core parameters are held constant (Table \ref{tabII}). Furthermore, equal electron and ion temperatures and $Z_\text{eff}$ of 2 are assumed. Fixing these parameters reduces the dimensionality of the problem. 
The separatrix temperature is also fixed at 50 eV, consistent with Spitzer-Härm conduction limited scrape-off layer for MAST-U at the operated power level. Future extensions of this work will relax these constraints.

\begin{table}[h!]
\centering
\caption{Fixed parameters used in all simulations.}\label{tabII}
\begin{tabular}{l l c}
\hline
\textbf{Parameter} & \textbf{Description} & \textbf{Value} \\
\hline
 & Equilibrium shape (64-term Fourier series) & fixed \\
$a$                 & Minor radius                         & 0.55 m \\
$R$                 & Major radius                         & 0.85 m \\
$\beta_{\text{N}}$  & Normalized core beta                 & 2.65 \\
$I_p$               & Plasma current                       & 740 kA \\
$B_\phi$            & Toroidal magnetic field              & 0.52 T \\
$Z_{\mathrm{eff}}$  & Effective charge                     & 2.0 \\
$Z_i$               & Main ion charge                      & 1.0 \\
$Z_{\mathrm{imp}}$  & Impurity charge                      & 6.0 \\
$T_{e,\text{sep}}$ & Electron temperature at the separatrix & 50 eV \\
$T_i = T_e$         & Ion-electron temperature equality    &  \\
\hline
\end{tabular}
\end{table}

\subsection{Generation of self-consistent Grad-shafranov equilibria}

The generation of self-consistent equilibria across a sampled parameter space has been applied in previous studies to construct physically consistent training datasets \cite{Bruncrona2025, KitEPS2025}. In the present implementation, for each sampled combination of pedestal parameters, smooth and
continuous temperature and density profiles are constructed using a modified hyperbolic tangent ($m$-tanh) parameterisation \cite{Groebner2001}.

Each sampled profile pair, together with the fixed plasma shape and global parameters corresponding to MAST-U discharge 49108, is passed to HELENA \cite{Huysmans1991HELENA}, which computes the equilibrium including the self-consistent bootstrap current, which can dominate the pedestal current profile due to steep pressure gradients. The Redl bootstrap current model was selected in this study \cite{Redl2021}. The inductively driven current is assumed to be fully diffused and follows the plasma conductivity profile. Its amplitude is adjusted to give the total current specified in the input. 

For each case, HELENA is iteratively run, adjusting the the core slope of the temperature profile until a target confined plasma normalized beta, $\beta_N$, is achieved. Following this procedure, 128 self-consistent equilibria simulations were initialized and 123 of them converged.

To prepare the equilibrium data from HELENA for gyrokinetic analysis, the output is post-processed using CHEASE, which is used primarily to smoothly map the magnetic flux surfaces to a R–Z grid and to generate a mapping file, which provides magnetic geometry input for GENE \cite{Lutjens1996CHEASE}. Alongside this, a profile input file is constructed to specify the plasma profiles on a $\rho$ grid. For the inputs, rotation values are estimated using the radial force balance \cite{Landreman2012}.

\subsection{Local linear GENE simulations}

To capture the radial variation of stability across the pedestal, six radial locations are selected for each equilibrium. Although the local flux-tube approximation is not expected to valid across the entire pedestal width, recent modeling studies have shown that local linear gyrokinetic calculations can still provide useful stability and transport information for reduced modeling workflows \cite{Li2026NSTX,Hatch2026DIIID}. In this proof-of-principle work, this limitation is therefore considered acceptable, and model extensions with global pedestal simulations will be pursued in future studies. 

The selected positions are uniformly spaced from the pedestal top of the broader profile (either temperature or density) to a location corresponding to 95\% of the pedestal width, which is around $\rho \sim 0.99$ with the exact value depending on the pedestal width. Here, $\rho$ denotes the normalized radial coordinate. At each radial location, simulations are carried out over a set of eleven normalized binormal wavenumbers on the ion-scale,
$k_y \rho_s$, spanning the range $0.1 \le k_y \rho_s \le 1.0$, with an additional point at $k_y=0.05$. This range resolves the relevant ion-scale instabilities expected in the pedestal region. For each $(x_0, k_y)$ combination, GENE computes linear instability properties, including the growth rate, frequency, and species-resolved particle and heat fluxes. 
The full set of simulations corresponds to 123 equilibria, six radial locations, and eleven wavenumbers per equilibrium. After removing non-converged simulations and cases with unphysical target quantities, the final dataset contains approximately 7500 local linear gyrokinetic simulations.

From the linear GENE simulations, we extract the species-resolved particle and heat fluxes. The total particle flux is taken as the sum of electrostatic and electromagnetic contributions.

\begin{equation}
  \Gamma_s = \Gamma^{\mathrm{ES}}_s + \Gamma^{\mathrm{EM}}_s,
  \qquad
  Q_s = Q^{\mathrm{ES}}_s + Q^{\mathrm{EM}}_s
  - \frac{3}{2}\,T_s\,\Gamma_s.
\end{equation}
where $T_s$ is the local equilibrium temperature.
The corresponding particle and heat diffusivities are defined in terms of the equilibrium gradients, and expressed using the normalized gradient scale lengths $a/L_{n,s}$ and $a/L_{T,s}$ as
\begin{equation}
  D_s = \frac{\Gamma_s}{(a/L_{n,s})\,n_s},
  \qquad
  \chi_s = \frac{Q_s}{(a/L_{T,s})\,n_s\,T_s}.
\end{equation}

Linear simulations do not include nonlinear saturation mechanisms and therefore do not provide reliable absolute flux levels. Ratios of diffusivities are more useful as transport fingerprints, since they describe the relative contribution of particle, electron heat, and ion heat transport channels \cite{Kotschenreuther2019}. In particular, $\chi_i/\chi_e$ and $D_e/\chi_e$ can be used to distinguish between different instability regimes \cite{Kotschenreuther2019}. The diffusivity ratios $\chi_i/\chi_e$ and $D_e/\chi_e$, together with the linear growth rate and real frequency, are therefore used as target quantities for the surrogate model.

\section{\label{sec:surrogate}Surrogate model} 

\begin{figure*}[t]
    \centering
    \includegraphics[width=\textwidth]{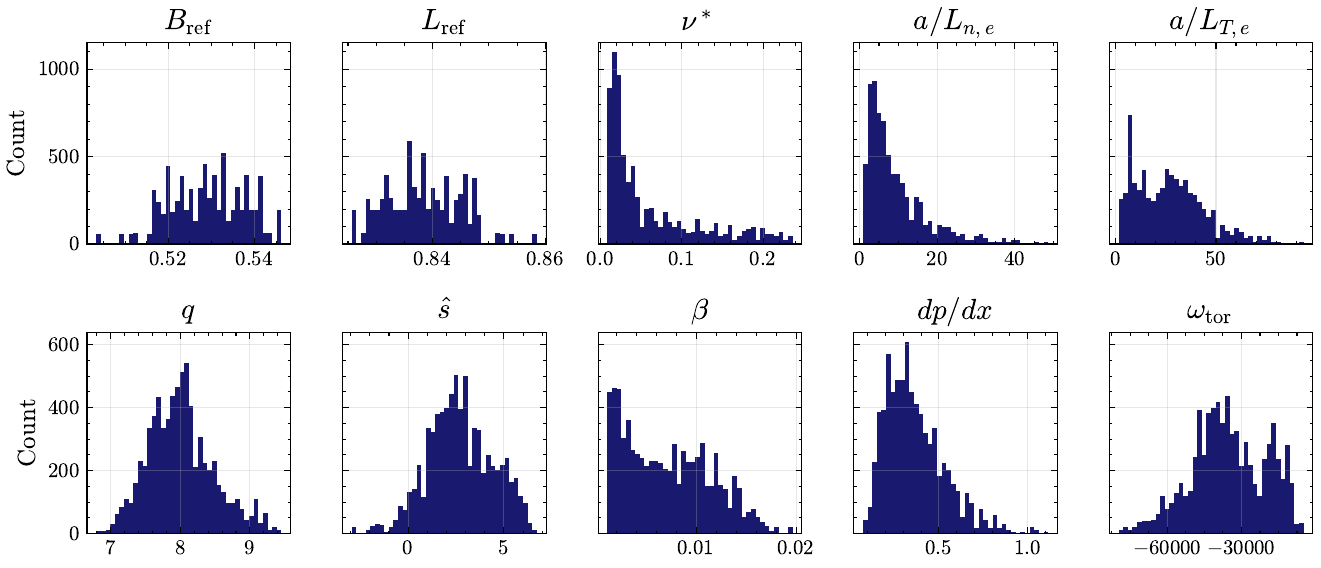}
    \caption{Distributions of the input parameters used to train the surrogate model.}
    \label{fig:input_distributions}
\end{figure*}

\subsection{Inputs and targets}

The surrogate inputs are selected to describe the local plasma conditions relevant for linear gyrokinetic stability in GENE for this reference MAST-U discharge (Fig. \ref{fig:input_distributions}). Each training sample corresponds to one local simulation at a radial location $x_0$ and binormal wavenumber $k_y$. The radial dependence enters through the local equilibrium quantities evaluated at $x_0$, while the normalized wavenumber $k_y\rho_s$ is included explicitly as an input feature.

The primary drivers of microinstabilities are the normalized density and temperature gradients, $a/L_{n_e}$ and $a/L_{T_e}$. Since ion and electron temperatures are assumed equal in the present dataset, the electron temperature gradient also sets the ion temperature gradient used in the simulations.
In addition to these, the normalized pressure gradient is included as a measure of the total pressure drive, which becomes particularly important in the electromagnetic regime and for instabilities such as KBMs. 

Magnetic geometry is represented through the safety factor $q$ and magnetic shear $\hat{s}$, which determine the structure of magnetic field line topology and influence both the stability thresholds and spatial characteristics of the modes. These parameters enter directly into the gyrokinetic equations and are important for capturing geometric effects on turbulence.

Collisionality affects electron dynamics and setting the balance between different instability regimes, for example between ITG and TEM. The plasma beta governs the importance of electromagnetic fluctuations and is a critical parameter for the onset of KBM and other electromagnetic  modes. In addition, the local toroidal rotation frequency $\omega_{tor}$, estimated from radial force balance, is included as an approximate representation of equilibrium flow effects related to the radial electric field, in anticipation of future global extension of the model that can apply the flow shear stabilisation mechanisms.

In addition to these physically motivated quantities, the reference magnetic field $B_{ref}$ and reference length scale $L_{ref}$
are included to ensure consistency with the normalization used in the gyrokinetic simulations and in derived quantities such as $\beta$ and collisionality. However, because the dataset is constructed around a single MAST-U reference discharge, with variation introduced primarily through the sampled temperature and density profiles, these reference quantities exhibit only limited variation across the dataset.

\begin{figure}[t]
    \centering
    \includegraphics[width=\columnwidth]{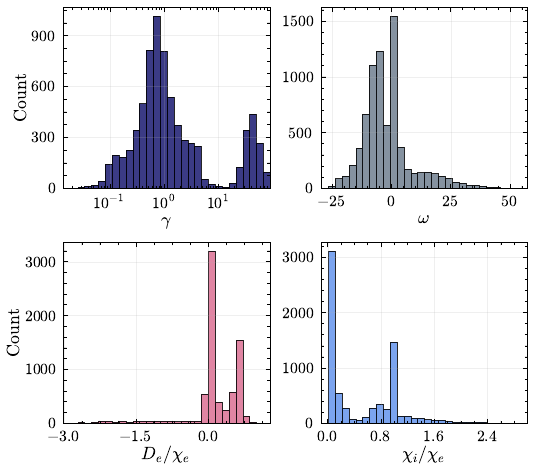}
    \caption{Distributions of the local linear GENE simulation outputs used as target variables for the surrogate model. The growth rate $\gamma$ and real frequency $\omega$ are given in GENE-normalized frequency units, i.e. normalized to $c_\mathrm{ref}/L_\mathrm{ref}$.}
    \label{fig:target_distributions}
\end{figure}

The target quantities are the growth rate $\gamma$, real frequency $\omega$, and diffusivity ratios $D_e/\chi_e$ and $\chi_i/\chi_e$, given in the normalized units used by the corresponding linear GENE simulations (Fig. \ref{fig:target_distributions}).

\subsection{Training setup}

All data points from the same HELENA sample are assigned to the same subset (training, validation or test), which guarantees that no equilibrium contributes simultaneously to multiple subsets. The training set is used for training the model, the validation set for evaluating model generalization within an unseen dataset during training, and test set for eventual evaluation of the trained model generalization. The training, validation, and test sets are constructed to target fractions of 70\%, 10\%, and 20\% of the total dataset, respectively; however, the exact proportions are approximate, as they are constrained by the requirement that all data associated with a given equilibrium are assigned to the same subset.
For the present dataset, this results in 86 equilibrium profiles used for training, 12 for validation, and 25 held out for testing, corresponding to approximately 5,200, 800, and 1,600 linear GENE simulations, respectively.

The distribution of the training inputs used in the ML model, are shown in Figure \ref{fig:input_distributions}.  The input values are shown in the normalized units used by GENE for the corresponding linear gyrokinetic simulations.

Prior to training, all input features are further standardized using z-score normalization, by subtracting the training-set mean and dividing by the training-set standard deviation. This ensures comparable numerical scales across variables and improves the stability and convergence of the optimization process.

\subsection{Model architectures}

\begin{figure*}[t]
    \centering
    \includegraphics[width=\textwidth]{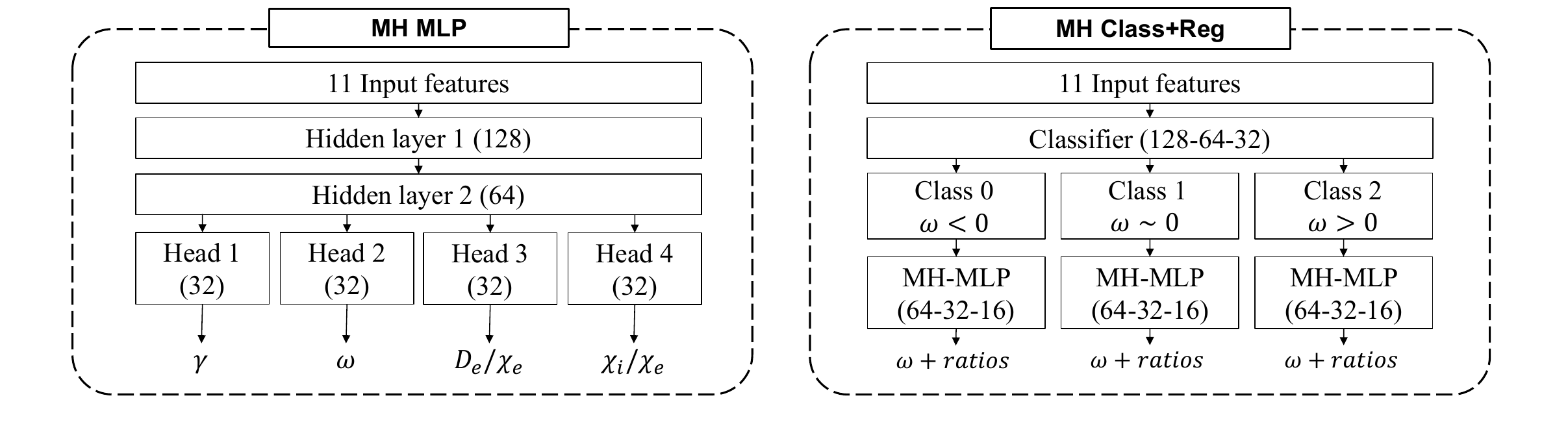}
    \caption{
    Diagrams of the surrogate-model architectures. The direct multi-head multilayer perceptron (MH-MLP) uses shared hidden layers followed by separate output heads for each target quantity. The multi-head classification--regression model (MH Class--Reg) first assigns each sample to a frequency-based class, after which a class-specific multi-head regressor predicts the clustered targets.
    }
    \label{fig:model_architectures}
\end{figure*}

\subsubsection{Multi-head Multilayer Perceptron (MH-MLP)}

The first model considered is a multi-head multilayer perceptron (MH-MLP) which predicts all target quantities from the same set of local input quantities \cite{Caruana1997}. Each training sample corresponds to a single pair of radial position and perpendicular wavenumber, $(x_0, k_y)$, within a given equilibrium.

A multi-head architecture is used because the target quantities are physically related outputs of the same linear gyrokinetic calculation. The growth rate, frequency, and transport ratios are all determined by the same underlying instability, but they can exhibit different functional behavior across parameter space. In particular, the growth rate often varies more smoothly, while the frequency and diffusivity ratios can show sharper transitions associated with changes in instability regime.

The model consists of a shared fully connected network followed by separate output heads for each target quantity. The shared layers learn a common representation, while the output heads allow the model to specialize the prediction for each target (Fig. \ref{fig:model_architectures}). 
The shared network contains two hidden layers with 128 and 64 neurons. Each output head contains an additional hidden layer with 32 neurons followed by a scalar output layer. The model is trained using the Adam optimizer with a mean squared error (MSE) loss, with the total loss formed from the regression losses of the individual target heads. Dropout regularization is applied during training to reduce overfitting.

\subsubsection{\label{sec:classreg}Multi-head Classification--Regression model (MH Class-Reg)}

The second model considered is a multi-head classification--regression model to capture regime-dependent structure in the target quantities. This is particularly relevant for the frequency and diffusivity ratios, which can exhibit sharp transitions associated with changes in the dominant instability mode. A single global regression model can therefore smooth across these transitions and produce intermediate values that are not representative of any mode.

The model first assigns each sample to one of three frequency-based classes, with values corresponding to $\omega < -1$, $-1 \leq \omega \leq 1$, and $\omega > 1$. The frequency is used for the classification step because it provides an indicator of changes in mode propagation direction and separates the data into broad regimes without imposing a full mode classification. This avoids defining classes from multiple simultaneous transport-ratio thresholds, which can be overly restrictive for the present dataset.

After classification, the input is passed to a class-specific multi-head regression network. Each class-specific regressor predicts $\omega$, $D_e/\chi_e$, and $\chi_i/\chi_e$. Thus, the model consists of one classifier and three separate multi-head regressors, one for each frequency class (Fig. \ref{fig:model_architectures}).

The classifier and regressors are implemented as fully connected neural networks. The classifier contains hidden layers with 128, 64, and 32 neurons and is trained using a cross-entropy loss. Each class-specific regressor contains shared hidden layers with 64 and 32 neurons, followed by separate target-specific output heads with 16 neurons. The regressors are trained using a mean squared error loss. A dropout rate of 0.1 is applied during training, and input features are standardized separately for the classification and regression stages.

This formulation preserves joint prediction of related target quantities while allowing different regression mappings to be learned in different frequency regimes. It provides a less restrictive alternative to a hard mode classification, since the frequency-based classes capture broad changes in propagation direction without requiring every point to satisfy a fixed combination of transport-ratio thresholds.

\section{Results}
The surrogate models are evaluated using three complementary tests. First, pointwise performance is assessed on a standard test set consisting of unseen equilibria from the sampled parameter space. Second, structured generalization tests are performed using beta scans and dense $(x_0,k_y)$ grids, which test whether the models reproduce physically meaningful trends across electromagnetic drive, radial position, and wavenumber. Finally, the predicted quantities are used to identify KBM-like regimes and compared with classifications derived from the GENE results.

\subsection{Multi-head MLP performance}

\begin{figure*}[t]
    \centering
    \includegraphics[width=\textwidth]{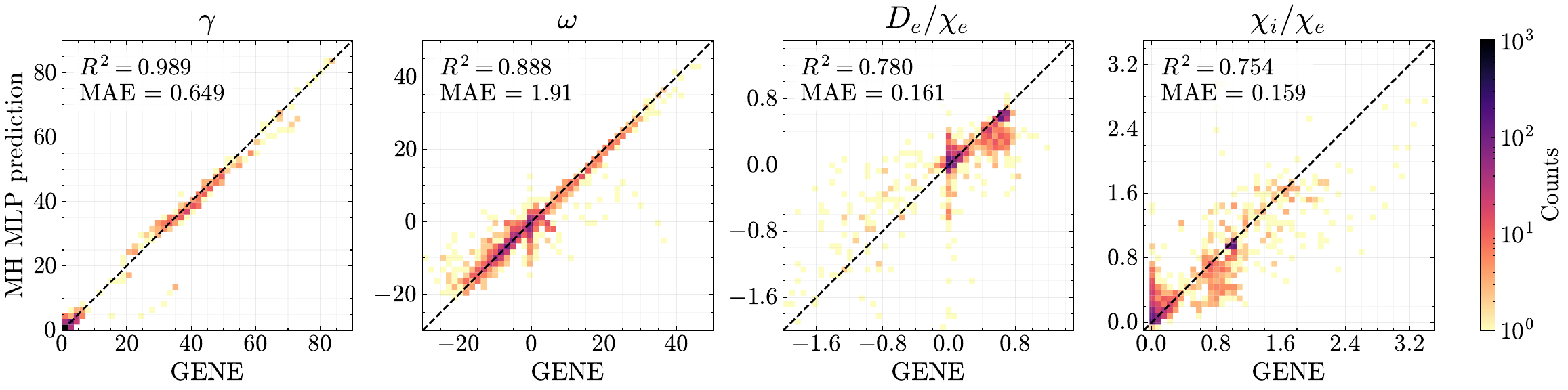}
    \caption{
    Parity plots comparing predictions from the multi-head multilayer perceptron (MH MLP) model with linear gyrokinetic simulations for the test set. Shown are the growth rate $\gamma$, frequency $\omega$, and the diffusivity ratios $D_e/\chi_e$ and $\chi_i/\chi_e$. The dashed line indicates perfect agreement. 
    }
    \label{fig:mlp_parity}
\end{figure*}

The direct MH-MLP reproduces the growth rate most accurately among the four target quantities (Fig. \ref{fig:mlp_parity}). The frequency is also captured reasonably well, although larger scatter is observed for negative-frequency modes and near the transition between positive and negative frequency.

The diffusivity ratios are more challenging (Fig. \ref{fig:mlp_parity}). The two-dimensional histograms show that the GENE values of $D_e/\chi_e$ and $\chi_i/\chi_e$ cluster into distinct regions, while the MH-MLP predictions are more diffuse. The model reproduces the approximate range of the ratios, but it does not fully preserve the clustered structure of the target data.

These results motivate the use of the classification--regression model described in Sec.~\ref{sec:classreg}, which is applied to $\omega$ and the diffusivity ratios.

\subsection{Classification--regression performance}

The classification--regression model is evaluated for the clustered targets, namely the frequency $\omega$ and the diffusivity ratios $D_e/\chi_e$ and $\chi_i/\chi_e$. Compared with the direct MH-MLP, this approach reduces the mean absolute error for these targets and produces sharper transitions in structured scans. The improvement is therefore most evident in the beta and dense $(x_0,k_y)$ evaluations discussed below, where the classification--regression model better preserves regime-like changes in the predicted quantities.

The $R^2$ values do not improve systematically and can be lower than those obtained with the direct MH-MLP (Table \ref{tab:clustered_target_metrics}. This reflects the sensitivity of the full pipeline to classification errors. When a point is assigned to the wrong frequency regime, the subsequent class-specific regressor can produce a large pointwise error. Such errors occur most often near regime boundaries.

The frequency-based classifier assigns the correct class for 92\% of the test samples. To separate the performance of the class-specific regressors from the performance of the classifier, an oracle test is also performed. In this test, the true frequency class is used to select the regressor. The oracle results show substantially lower mean absolute error (MAE) and higher $R^2$ for all three targets, when compared to the models. This indicates that the class-specific regressors are able to model the targets more accurately within each regime, while the main limitation of the full pipeline is the robustness of the classification step.

Overall, these results suggest that regime-conditioned regression improves the prediction of clustered targets in terms of typical error, while sharp mode transitions remain the main challenge of the model.

\begin{table}[htbp]
\centering
\caption{
Test-set performance for the clustered targets. The oracle class--regression model uses the true frequency class to select the class-specific regressor, showing the performance achievable when the regime is identified correctly.
}
\label{tab:clustered_target_metrics}
\setlength{\tabcolsep}{4.5pt}
\renewcommand{\arraystretch}{1.15}
\begin{tabular}{lcccccc}
\toprule
& \multicolumn{2}{c}{$\omega$}
& \multicolumn{2}{c}{$D_e/\chi_e$}
& \multicolumn{2}{c}{$\chi_i/\chi_e$} \\
\cmidrule(lr){2-3}
\cmidrule(lr){4-5}
\cmidrule(lr){6-7}
Model
& MAE & $R^2$
& MAE & $R^2$
& MAE & $R^2$ \\
\midrule
MH-MLP
& 1.91 & 0.888
& 0.161 & 0.780
& 0.159 & 0.754 \\
MH Class--Reg
& 1.33 & 0.866
& 0.124 & 0.728
& 0.117 & 0.756 \\
Oracle
& 0.77 & 0.965
& 0.074 & 0.923
& 0.069 & 0.908 \\
\bottomrule
\end{tabular}
\end{table}

\subsection{Generalization tests}
\subsubsection{Beta scan}
\begin{figure}[htbp]
    \centering
    \includegraphics[width=\columnwidth]{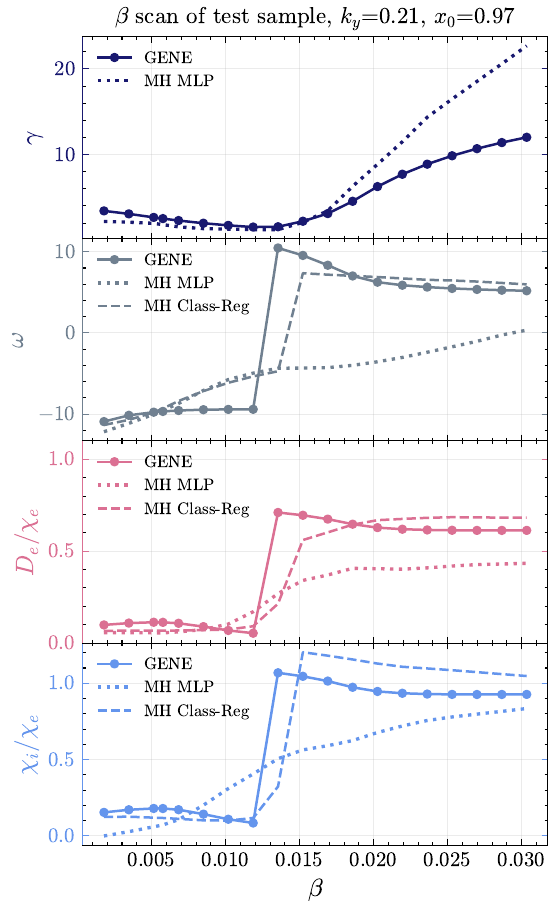}
    \caption{Comparison of surrogate model predictions and linear GENE simulations for a representative beta scan at fixed $(x_0, k_y)$. Shown are the growth rate $\gamma$, frequency $\omega$, and the diffusivity ratios $D_e/\chi_e$ and $\chi_i/\chi_e$ as functions of plasma beta.}
    \label{fig:beta_scan}
\end{figure}

To assess generalization with respect to electromagnetic effects, a scan in plasma beta is performed for an unseen equilibrium (Fig. \ref{fig:beta_scan}). The models are evaluated at fixed $(x_0, k_y)$ locations with $k_y \rho_s \lesssim 0.5$, where electromagnetic effects and KBM-like behavior are expected to become important.

The surrogate model reproduces the overall trends of the simulations across the beta range. As $\beta$ increases, the growth rate increases and the mode frequency shifts towards the ion diamagnetic direction. The diffusivity ratios also shift towards values expected for KBM-like modes, namely $D_e/\chi_e \sim 2/3 $ and $\chi_i/\chi_e \sim 1$, as discussed further in Sec \ref{sec:kbm_classification}.

The classification-regression formulation improves the representation of sharp transitions in the diffusivity ratios and frequency compared to the MH-MLP model, which tends to smooth these transitions over a broader range of $\beta$ (Fig. \ref{fig:beta_scan}). The surrogate model does not reproduce the exact location of the mode transition in beta space, although the transition is captured approximately through changes in the frequency, and diffusivity ratios (Fig. \ref{fig:beta_scan}). The MH-MLP model captures the trend in the growth rate, however at higher beta values, the model tends to overpredict the increase in growth rate relative to the GENE simulations (Fig. \ref{fig:beta_scan}). This behavior likely reflects the limited sampling of strongly electromagnetic regimes within the training dataset.

\subsubsection{$k_y, x_0$ scan}

\begin{figure*}[t]
    \centering
    \includegraphics[width=\textwidth]{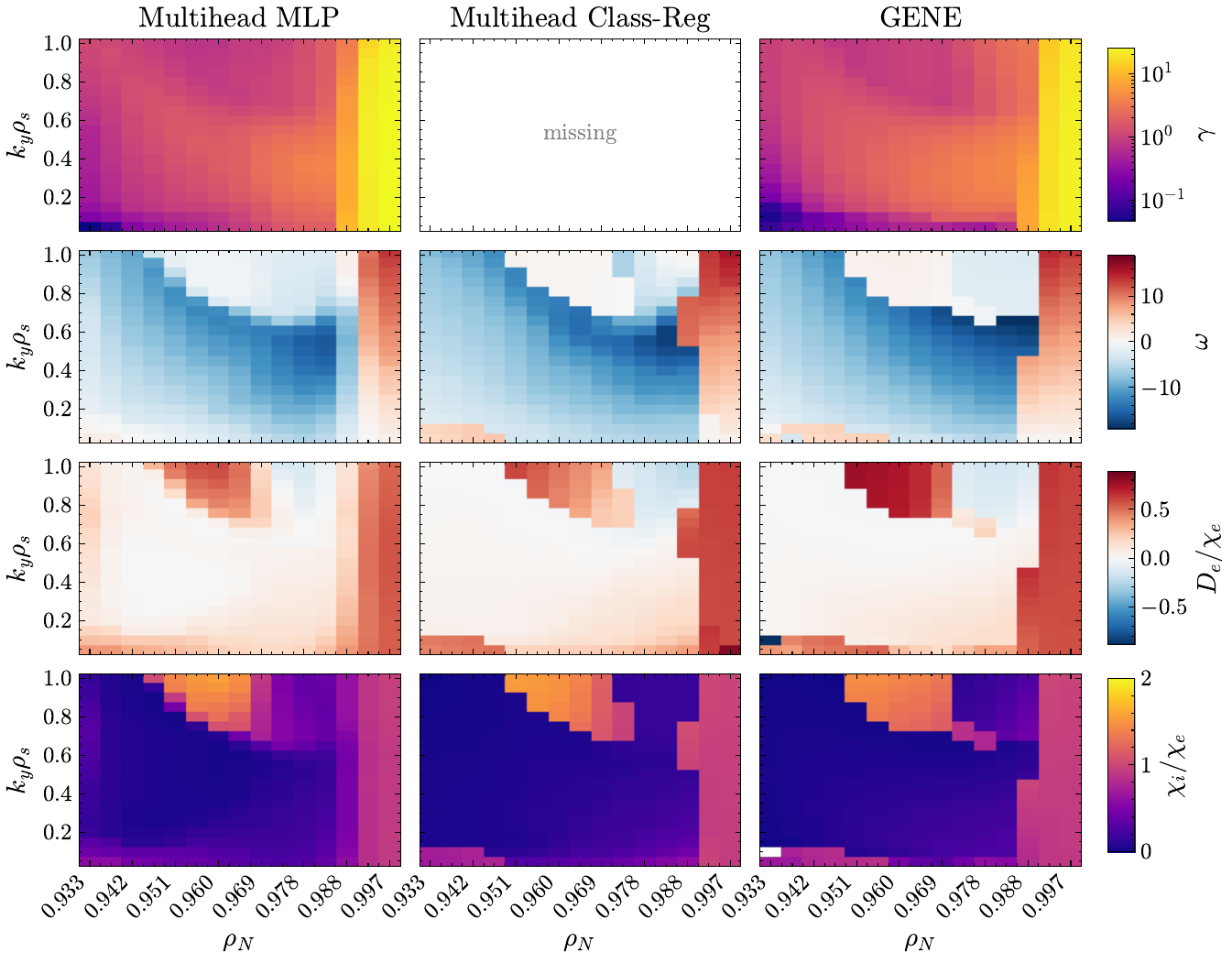}
    \caption{Comparison of surrogate model predictions and linear GENE simulations across a $(x_0,k_y)$ grid in the pedestal for an unseen equilibrium. Shown are the growth rate $\gamma$, frequency $\omega$, and diffusivity ratios as functions of $\rho_N$ and $k_y\rho_s$.The surrogate model reproduces the overall instability structure and captures sharp transitions between different transport regimes.}
    \label{fig:x0_ky_scan}
\end{figure*}

To assess the interpolation performance of the surrogate model, a higher-resolution evaluation in $(x_0, k_y)$ space is performed using an unseen equilibrium (Fig. \ref{fig:x0_ky_scan}). Linear gyrokinetic simulations are carried out at 15 uniformly spaced radial locations in the pedestal, with 20 wavenumber values spanning the range $0.05 \le k_y \rho_s \le 1.0$, providing a denser sampling than used in training. 

The surrogate models reproduce the overall structure of the instability spectrum, including the radial increase in growth rate towards the separatrix and the variation of mode properties across the wavenumber space. 

Although the precise locations of the transition do not always coincide exactly with the GENE results, the classification-regression approach improves the representation of mode transitions in comparison to the MH-MLP model. The predicted diffusivity ratios and frequency exhibit sharper transitions and more clearly defined regions associated with different instability fingerprints.

\subsection{\label{sec:kbm_classification}KBM classification}

To test whether the surrogate preserves physically consistent relationships between the predicted outputs, the predicted quantities are used to identify KBM-like modes and compared with labels obtained from the GENE results. KBM identification is of particular interest because KBM onset is widely used as a physics basis for pedestal-width predictions in reduced models such as EPED. The classification is based on the transport-fingerprint approach of Kotschenreuther et al. \cite{Kotschenreuther2019}, in which KBM-like modes are expected to have diffusivity ratios close to $D_e/\chi_e \simeq 2/3$ and $\chi_i/\chi_e \simeq 1$, together with propagation in the ion diamagnetic direction. In this work, a point is labeled KBM-like if $0.5 \leq D_e/\chi_e \leq 0.7$, $0.9 \leq \chi_i/\chi_e \leq 1.1$, and the frequency is consistent with propagation in the ion diamagnetic direction.

This classification should be interpreted as an approximate identification of KBM-like behavior rather than a definitive mode identification. A more complete classification would require additional diagnostics, such as parity analysis or dedicated beta scans. Nevertheless, it provides a test of whether the surrogate model preserves physically meaningful and consistent instability fingerprints.

Using these criteria, the MH Class--Reg model gives an overall test-set accuracy of 0.98 and a macro-F1 score of 0.97, where the macro-F1 is the unweighted mean of the class-wise F1-scores. The corresponding values for the MH-MLP are 0.97 and 0.95, respectively (Table \ref{tab:kbm_classification}). For the KBM-like class, recall increases from 0.86 to 0.91 with the MH Class--Reg model, indicating that a larger fraction of the true KBM-like points are recovered. The KBM-like F1-score also increases from 0.91 to 0.95, while the non-KBM class remains accurately identified by both models.

The reduced accuracy for KBM-like modes is expected, since the dataset is imbalanced, with substantially fewer KBM-like samples than non-KBM points in the test set (Table \ref{tab:kbm_classification}). In addition, the diffusivity-ratio thresholds are inherently approximate. Overall, the results in Table~\ref{tab:kbm_classification} indicate that the surrogate predictions retain sufficient physical structure to accurately predict the dominant KBM-like regions of the test set.

\begin{table}[htpb]
\centering
\caption{
Class-wise KBM-like classification performance on the test set. Labels are derived by applying the transport-fingerprint criteria to the GENE outputs and to the surrogate predictions. Precision is the fraction of samples predicted to belong to a class that are correctly classified, recall is the fraction of true samples of that class that are recovered, and F1 is their harmonic mean. Support denotes the number of true samples in each class.
}
\label{tab:kbm_classification}
\setlength{\tabcolsep}{5pt}
\renewcommand{\arraystretch}{1.15}
\begin{tabular}{llcccc}
\toprule
Model & Class & Precision & Recall & F1 & Support \\
\midrule
MH-MLP & non-KBM & 0.97 & 0.99 & 0.98 & 1308 \\
& KBM & 0.97 & 0.86 & 0.91 & 300 \\
\midrule
MH Class--Reg & non-KBM & 0.98 & 1.00 & 0.99 & 1308 \\
& KBM & 0.98 & 0.91 & 0.95 & 300 \\
\bottomrule
\end{tabular}
\end{table}

\section{Discussion and conclusions}

This work investigated machine-learning surrogate models for local linear gyrokinetic simulations in the pedestal region. The motivation is to reduce the cost of obtaining linear gyrokinetic quantities needed in quasilinear and reduced transport modeling, while retaining a closer connection to kinetic microinstability physics than reduced descriptions.

A dataset of local linear GENE simulations was generated within an experimentally motivated MAST-U pedestal parameter space. Sampled pedestal temperature and density profiles were used to construct self-consistent equilibria, and local linear simulations were performed across multiple radial locations and ion-scale binormal wavenumbers. From these simulations, growth rates, real frequencies, and diffusivity-ratio fingerprints were extracted as surrogate-model targets.

Two neural-network approaches were evaluated. A multi-head MLP was used to predict all target quantities from local plasma inputs, allowing shared information between related gyrokinetic outputs. A multi-head classification--regression model was then introduced for the more regime-dependent targets, using a frequency-based classification step followed by class-specific regressors for the frequency and diffusivity ratios.

The direct MH-MLP accurately reproduced the growth rate and gave reasonable predictions for the real frequency, but the diffusivity ratios were more challenging, likely due to the clustered structure and sharp transitions associated with changes in instability regime. The MH Class--Reg model reduced the mean absolute error for the clustered targets and produced sharper transitions in structured scans, although classification errors near mode-transition regions limited the improvement in global $R^2$. However, the level of predictive accuracy required in practice will ultimately depend on the intended transport-model application and remains to be established in future studies.

The oracle classification--regression results help to separate the regression performance from the classification performance: When the correct frequency class is used to select the regressor, the prediction accuracy improves substantially for all clustered targets. This suggests that the class-specific regressors are able to learn the relevant mappings within each regime, and that the main limitation of the full pipeline is the regime-identification step.

The classification stage therefore remains an area of further improvement. In the present work, a simple frequency-based classification was used to define three broad regimes. This provides a useful proof of principle, but it does not fully capture all mode transitions, particularly cases where transport fingerprints change within the same frequency class. Future work could explore more detailed mode classifications, including additional classes associated with different instability types. Such an extension would likely require a larger and more balanced dataset to ensure that all regimes are sufficiently represented.

A probabilistic treatment of the classification step may also be valuable. The class probabilities produced by the classifier could be used as an uncertainty indicator, especially near transition regions where the maximum class probability is low. Rather than selecting a single regressor, future models could combine class-specific predictions using the predicted class probabilities.

Several limitations of the present study should be considered when interpreting the surrogate-model performance and its applicability beyond the current dataset. First, the dataset is constructed from local linear gyrokinetic simulations restricted to ion-scale wavenumbers. The surrogate is therefore trained to reproduce linear instability properties and associated transport fingerprints, rather than fully saturated turbulent fluxes, and does not include electron-scale or multiscale turbulence. In addition, the model inherits the assumptions of the local flux-tube simulations, including the assumption of scale separation between equilibrium profiles and turbulent fluctuations, which may become less accurate in the steep-gradient pedestal region.

A further limitation arises from the restricted parameter space considered in this work. The dataset is based on a single MAST-U discharge with several equilibrium and machine parameters held fixed, including plasma shape, magnetic field, plasma current, and species composition. While this enables a controlled proof-of-principle study focused on pedestal profile variation, it limits the ability of the surrogate model to generalize beyond the sampled conditions.

The present work demonstrates that surrogate models can reproduce several outputs of local linear gyrokinetic pedestal simulations, but several extensions are needed before the approach can be applied more broadly. Future work will expand the sampled operational space beyond a single discharge, enabling the training of surrogate models across a broader range of MAST-U operational regimes.

\begin{acknowledgments}

This work has been carried out within the framework of the EUROfusion Consortium, funded by the European Union via the Euratom Research and Training Programme (Grant Agreement No 101052200 —EUROfusion). Views and opinions expressed are, however, those of the author(s) only and do not necessarily reflect those of the European Union or the European Commission. Neither the European Union nor the European Commission can be held responsible for them. The work of Aaro Järvinen, Adam Kit, and Amanda Bruncrona has been partially supported by the Research Council of Finland grant no. 355460, and the work of Daniel Jordan and Anna Niemelä by the Research Council of Finland grant no. 358941. The authors wish to acknowledge CSC – IT Center for Science, Finland, for computational resources.
\end{acknowledgments}

\section*{Data Availability Statement}

The authors do not have permission to share the experimental data analyzed during the current study. Data access can be obtained
by creating a user account for the MAST-U computer system.

The generated dataset used to train and evaluate the surrogate models is openly available in Zenodo at https://doi.org/10.5281/zenodo.21937872. \cite{Niemela2026Zenodo}

The software tools and trained neural-network models developed in this study are available in the GENE-NN repository at https://github.com/DIGIfusion/GENE-NN. The version corresponding to the results presented here is archived as v0.1.0.


\section*{References}
\bibliography{aipsamp}

\end{document}